\documentclass[reprint, longbibliography, superscriptaddress,amsmath,amssymb,aps, prl]{revtex4-2}

\usepackage{graphicx} 
\usepackage{bm}
\usepackage{xr-hyper}

\usepackage{dcolumn}
\usepackage{nicematrix}
\usepackage{placeins} 

\newcommand{\DU}[1]{\ensuremath{\ket{\downarrow \uparrow}}}

\date{\today}

\begin{document}

\title{Direct measurement of the energy spectrum of a quantum dot qubit}


\author{J. Reily}
\affiliation{Department of Physics, University of Wisconsin-Madison, Madison WI, 53706, United States}
\affiliation{Q-NEXT, Argonne National Laboratory, Lemont IL, 60439, United States}

\author{Daniel J. King}
\affiliation{Department of Physics, University of Wisconsin-Madison, Madison WI, 53706, United States}

\author{Jonathan C. Marcks}
\affiliation{Q-NEXT, Argonne National Laboratory, Lemont IL, 60439, United States}
\affiliation{Materials Science Division, Argonne National Laboratory, Lemont IL, 60439, United States}
\affiliation{Prizker School of Molecular Engineering, University of Chicago, Chicago IL, 60637, United States}

\author{M.A. Wolfe}
\affiliation{Department of Physics, University of Wisconsin-Madison, Madison WI, 53706, United States}

\author{Piotr Marciniec}
\affiliation{Department of Physics, University of Wisconsin-Madison, Madison WI, 53706, United States}

\author{E.S. Joseph}
\affiliation{Department of Physics, University of Wisconsin-Madison, Madison WI, 53706, United States}

\author{Tyler J. Kovach}
\affiliation{Department of Physics, University of Wisconsin-Madison, Madison WI, 53706, United States}

\author{Brighton X. Coe}
\affiliation{Department of Physics, University of Wisconsin-Madison, Madison WI, 53706, United States}




\author{Mark Friesen}
\affiliation{Department of Physics, University of Wisconsin-Madison, Madison WI, 53706, United States}

\author{Benjamin D. Woods}
\affiliation{Department of Physics, University of Wisconsin-Madison, Madison WI, 53706, United States}

\author{M.A. Eriksson}
\affiliation{Department of Physics, University of Wisconsin-Madison, Madison WI, 53706, United States}

\begin{abstract}

The mapping between gate voltages applied to a double quantum dot, and the parameters of a Hubbard-like Hamiltonian, is of utmost importance for understanding and operating spin qubits. State-of-the-art techniques for measuring Hamiltonian parameters (e.g., detuning axis pulsed spectroscopy, DAPS) provide details about energy levels; however, tunnel coupling estimates typically reveal only a small portion of the full Hamiltonian. Here, we demonstrate a Hamiltonian-agnostic technique for measuring the double dot energy spectrum over a wide energy range, at every value of the detuning, called delta-axis spectroscopy (DAXS).  We apply the DAXS method to obtain the energy spectrum of a Si/SiGe double quantum dot and use this data to extract the diagonal and off-diagonal couplings of a 15-level Hubbard-like Hamiltonian, demonstrating very good agreement with the experimental measurements.

\end{abstract}

\maketitle


Gate-defined semiconductor quantum devices are well-studied systems both as a platform for physics discovery~\cite{vanweesQuantizedConductancePoint1988, taruchaShellFillingSpin1996, livermoreCoulombBlockadeCoupled1996, goldhaber-gordonKondoEffectSingleelectron1998, vanderwielElectronTransportDouble2002, hensgensQuantumSimulationFermi2017} and for their potential to create large scale quantum computers using spin states in quantum dots~\cite{burkardSemiconductorSpinQubits2023, Neyens_2024, desmetHighfidelitySinglespinShuttling2025, ivlevOperatingSemiconductorQubits2025a, steinackerIndustrycompatibleSiliconSpinqubit2025}.  To coherently manipulate and readout the majority charge carriers confined in these devices, it is necessary to understand the Hamiltonian parameters of the system which often, if not always, depend on detuning.  Indeed, plotting calculated values of a qubit's energy as a function of detuning is often the most useful way to describe the qubit's states in $S$--$T_0$~\cite{pettaCoherentManipulationCoupled2005}, exchange-only~\cite{lairdCoherentSpinManipulation2010}, resonant exchange-only~\cite{medfordQuantumDotBasedResonantExchange2013}, Loss-Divincenzo (single-spin)~\cite{veldhorstTwoqubitLogicGate2015}, quantum dot hybrid~\cite{kimHighfidelityResonantGating2015}, single-hole spin~\cite{hendrickxSingleholeSpinQubit2020}, flopping mode spin~\cite{huFloppingmodeSpinQubit2023}, hopping spin~\cite{wangOperatingSemiconductorQuantum2024a}, and $S$--$T_-$ hole~\cite{zhangUniversalControlFour2025} qubit systems.  
Knowledge of the energy dispersion of excited states is also critical for avoiding Landau-Zener excitations to leakage states~\cite{miLandauZenerInterferometryValleyorbit2018} and for locating anticrossings where excited states can cause hotspots~\cite{blumhoff2022}.


Many common techniques exist to measure partial information about quantum dot Hamiltonian parameters.  Energy splittings can be measured using pulsed-gate spectroscopy (PGS)~\cite{elzermanExcitedstateSpectroscopy2004} and detuning axis pulsed spectroscopy (DAPS)~\cite{daps}.  Photon assisted tunneling (PAT)~\cite{wangChargeRelaxationSingleElectron2013} reveals low-lying, charge-like splittings and ground state couplings.  Landau-Zener-St\"uckelberg interferometry can extract tunnel couplings~\cite{miLandauZenerInterferometryValleyorbit2018}, although this may require complicated pulsing schemes, especially when probing couplings between excited states.  Linewidth broadening is a simpler technique that aims to measure ground state-to-ground state tunnel couplings~\cite{dicarlo}, although it does not account for the influence of nearby excited states~\cite{Zhao_Hu_2022}.  Other methods require resonators and elevated temperatures~\cite{Mi_Petta_2017, borjansProbingVariationIntervalley2021a}.  Most importantly, each of these techniques reveals only a subset of the parameters in the Hamiltonian matrix.

Here we demonstrate a new approach to obtain and analyze both diagonal and off-diagonal elements of the Hamiltonian matrix using only a simple square wave voltage pulse measurement.  We demonstrate that pulsing along the delta ($\delta$) axis (defined below) in gate voltage space directly maps out the energy levels of double-dot-based artificial molecules, including those used for semiconductor quantum dot qubits.  Because the pulse is along the $\delta$ direction, we refer to the technique as delta axis spectroscopy (DAXS).  By fitting these maps to the eigenvalues of a Hubbard Hamiltonian, both diagonal and off-diagonal matrix elements can be extracted, revealing the energy eigenstates and the effective tunnel couplings. DAXS can extract these parameters because it provides an experimental map of the energy dispersion as a function of detuning, rather than a single spectrum at a fixed detuning. For this reason DAXS provides substantially more information than PGS; nonetheless, we can still compare DAXS directly with PGS where the two have overlap, and we find good agreement. We report example extraction of energy levels and tunnel couplings, and we estimate the uncertainty in these parameters. Finally, we demonstrate a method to differentiate between energy levels of the double dot and features that arise from modulations in the density of states of the reservoirs, which commonly arise from the narrowing of the leads as they connect to quantum dots through quantum point contacts.

\begin{figure*}
\includegraphics{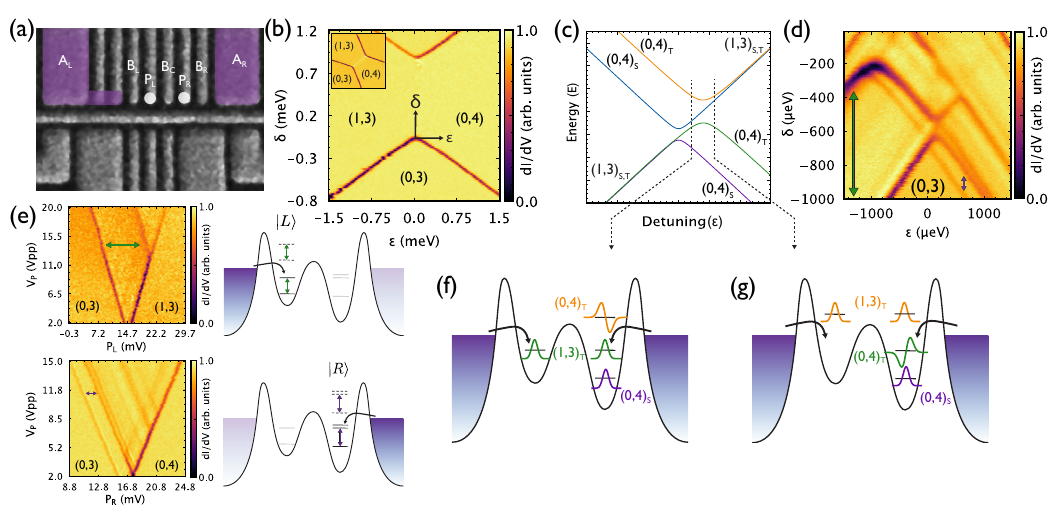}
\caption{\bf Overview of measurement regime and spectroscopy \rm (a) A false-colored scanning electron micrograph of a device lithographically identical to the one used in this experiment.  Gates are labeled P(B)[A] for Plunger(Barrier)[Accumulation]. (b) Stability diagram of the (1,3)-(0,4) charge configuration plotted as a function of $\varepsilon$ and $\delta$.  The inset shows the same charge regime as a function of the two plunger gate voltages.  (c) Energy dispersion corresponding to the (1,3)-(0,4) charge regime.  (d)  Average of four DAXS measurements of the double dot utilizing both reservoirs.  (e) PGS data for both $P_\text{L}$ (top) and $P_\text{R}$ (bottom) with corresponding double-dot schematics. The voltage pulsing, $V_\text{P}$, is represented with solid lines at the bottom of the pulse and dashed lines at the top, in units of peak to peak voltage.  Green and purple arrows depicting first visible splittings are shown here and also in (d). (f,g) Double-dot diagrams depicting schematic wavefunctions, following Ref.~\onlinecite{burkardSemiconductorSpinQubits2023}, on both the left (f) and right (g) sides of the excited state anticrossing shown in (c).}
\label{fig12}
\end{figure*}

\textit{DAXS}---We demonstrate DAXS on a double quantum dot (DQD) formed in a Si/SiGe Intel Tunnel Falls device~\cite{Neyens_2024, George_2025}, as shown in Fig.~\ref{fig12}(a). Dots are accumulated under plunger gates $P_\text{L}$ and $P_\text{R}$, barriers confining the dots are $B_\text{L}$, $B_\text{C}$, and $B_\text{R}$, and reservoirs are $A_\text{L}$ and $A_\text{R}$. The left reservoir is further extended under the two otherwise unused finger gates, as shown in the figure, and the lower half of the device forms the quantum dot charge sensor used throughout this work. We tune the double quantum dot to the (1,3)-(0,4) electron occupation regime, whose stability diagram is shown in Fig.~\ref{fig12}(b) with the two lowest energy electrons in the right dot forming a ``closed-shell'' singlet.  The stability diagram is measured along virtualized detuning $\varepsilon=\mu_{2}-\mu_{1}$ and delta $\delta=\frac{\mu_{2}+\mu_{1}}{2}$ axes (see main panel), where $\mu_1$ and $\mu_2$ refer to the chemical potentials of the left and right dot, respectively. These parameters are connected to gate voltages by energy calibration lever arms~\cite{vanderwielElectronTransportDouble2002}.  In these measurements, the color corresponds to the differential current through the charge sensor with respect to the oscillating voltage on both $P_\text{L}$ and $P_\text{R}$~\cite{SI}.


Figure~\ref{fig12}(c) shows an energy vs.\ detuning diagram of the type that is widely used to understand quantum dot phenomena. Here we include only the lowest energy singlet (S) and triplet (T) states for simplicity. Experimental measurement and visualization of this diagram---in both its qualitative and quantitative features---would enable extraction of both diagonal and off-diagonal elements of the Hamiltonian matrix.  Fig.~\ref{fig12}(d) shows the time-averaged lock-in measurement of the charge sensor current in the presence of a square wave voltage applied to gates $P_\text{L}$ and $P_\text{R}$, following the DAXS procedure described below. This illustrates the clear correspondence between DAXS results and the theoretical spectrum shown in Fig.~\ref{fig12}(c).  Many additional lines are visible in the DAXS plot of Fig.~\ref{fig12}(d), corresponding to excited states and excited-state anticrossings, features we discuss in detail below.  Additionally, Fig.~\ref{fig12}(d) is the average of four measurements taken at different reservoir ($A_\text{R}$) voltages to average out resonances from lead states, also discussed below.

To understand DAXS, it is useful to compare it with the PGS measurements shown in Fig.~\ref{fig12}(e).  PGS also employs a periodic square pulse applied to a quantum dot plunger gate.  As shown in the figure, the amplitude of the square wave is typically increased as the DC voltage on the plunger gate is varied, resulting in the characteristic V-shaped charge sensed measurement shown in Fig.~\ref{fig12}(e). Each additional line corresponds to an excited state entering the measurement window; here, the first orbital excited state splittings in dots L and R are indicated by the green and purple arrows, respectively.

The excited states in Fig.~\ref{fig12}(e) are also visible in the DAXS measurement of Fig.~\ref{fig12}(d), indicated by arrows of the same color.  In contrast to PGS, DAXS reveals the hybridization of these states and shows clearly the resulting anticrossings between the (1,3) and (0,4) states. Figure~\ref{fig12}(f,g) shows schematics of the DQD wave functions at two detunings, on either side of an excited state anticrossing, following the notation of Ref.~\onlinecite{burkardSemiconductorSpinQubits2023}. In DAXS, pulsed gate voltages are applied to both plunger gates simultaneously.  However, in contrast to PGS, the amplitude of these pulses are not swept.  The voltage pulses are applied in the $\delta$ direction (parallel to the polarization line).  In the vicinity of an anticrossing, pulsing along delta with tunneling possible into either quantum dot, allows for the loading of hybridized states, even as that hybridization changes across an anticrossing. If only one dot of the DQD is tunnel coupled to a source of carriers, accessing states on both sides of an anti-crossing would only be possible through co-tunneling processes~\cite{gustavssonDetectingSingleelectronTunneling2008, braakmanLongdistanceCoherentCoupling2013}.




\begin{figure}
\centering
\includegraphics{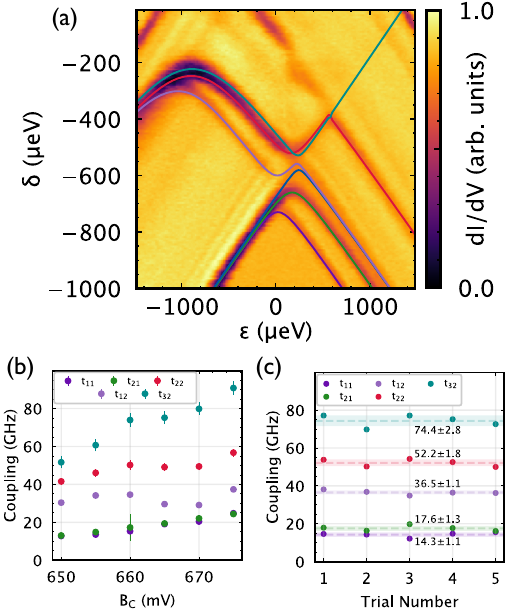}
\caption{\bf DAXS fitting and and extracted couplings.  \rm  (a) DAXS data overlaid with fits of eigenvalues from the 15x15 Hamiltonian, shown in Ref.~\onlinecite{SI} describing the levels and couplings.  (b) Plot of extracted couplings as a function of the center barrier gate voltage, $B_\text{C}$.  (c) Plot of extracted couplings over 5 different trials, showing scan-to-scan variability of a single tuning where $B_\text{C}=660$ $mV$.  One standard deviation is displayed as shaded regions for each coupling and the average is a dashed line.}
\label{fig3}
\end{figure}

Figure~\ref{fig3}(a) shows the average of four measurements like that shown in Fig.~\ref{fig12}(d) taken at different $A_\text{R}$ voltages. Overlaid are the eigenvalues of a Hamiltonian matrix consisting of DQD charge states and off-diagonal tunnel couplings.  This fit captures both the hybridized double dot states (straight-line segments) and their evolution as a function of detuning, visible in the curvature at anticrossings. The fitting process involves two steps: first, each peak corresponding to a double dot energy eigenstate is fit to a Lorentzian. We note that some peaks in a DAXS plot correspond to resonances from the reservoir and should not be fit; below we return to this point and explain how to identify and exclude those features. Second, the centers of these Lorentzians provide a set of points in the $\delta$ vs.\ $\varepsilon$ plane.  These points correspond to the eigenvalues of the Hamiltonian, and the Hamiltonian diagonal and off-diagonal terms are determined by a best fit to those points. Tunnel couplings $t_{ij}$ are extracted between combinations of ground and excited states, where $i$ ($j$) refers to the right (left) dot energy level. See Ref.~\onlinecite{SI} for the full fitting matrix and a detailed description of the fitting procedure.

It is immediately clear from both the raw DAXS data and the fits shown in Fig.~\ref{fig3}(a) that, in general, the tunnel couplings between higher lying excited states are larger than those between lower lying states. This general pattern is expected, due to tendency of higher-energy states states to be less localized, but it is not always followed: e.g., the red line in Fig.~\ref{fig3}(a) shows a sharp turn, nearly a kink, corresponding to a small tunnel coupling $t_{41}$. In fact, $t_{41}$ is small compared to the observed linewidth for the DAXS reservoir transitions, with this linewidth arising from the electron temperature of about 100 mK (2 GHz).

Figure~\ref{fig3}(b) also reports the extracted tunnel couplings for a range of $B_\text{C}$ gate voltages.  This gate is expected to control the tunnel coupling between the two dots, with more positive gate voltage typically corresponding to larger tunnel coupling. The five tunnel couplings reported in Fig.~\ref{fig3}(b) all have larger tunnel couplings for the most positive barrier gate voltage than they do for the most negative.  Indeed, many of the tunnel couplings increase monotonically over the range studied here.  Some nonmonotonic behavior is visible, for $t_{12}$ and $t_{22}$, and this may arise from the structure of the orbital states in each dot, which will be sensitive to atomic scale disorder in the heterostructure.
Fig.~\ref{fig3}(c) compares the extracted tunnel couplings for five repeated measurements with the same device gate voltages. The one-sigma variation in each of the tunnel couplings is reported, and these vary from 1.6 to 3.9~GHz. The error bars plotted in Fig.~\ref{fig3}(b) are the combination in quadrature of the variation shown in Fig.~\ref{fig3}(c) and the uncertainty arising in the global fit described above.

\textit{Distinguishing Quantum Dot States from Resonances in the Source Reservoirs}---The source for tunneling into the DQD in this work is the two reservoirs accumulated by gates $A_\text{L}$ and $A_\text{R}$, as shown in Fig.~\ref{fig12}(a). If these reservoirs were infinite in two dimensions, the density of states (DOS) would be constant as a function of energy and would produce no features in a DAXS plot. However, because of their finite size, the reservoirs exhibit quasi-1D effects and a non-uniform DOS whose effects are visible in transport experiments~\cite{limObservationSingleelectronRegime2009, mottonenProbeControlReservoir2010, escottResonantTunnellingFeatures2010,  TanKlimeckThesis} and also in PGS~\cite{daps}. These resonances also produce lines in DAXS plots, and it is important to distinguish between them and energy eigenstates of the DQD.

\begin{figure}
\centering
\includegraphics{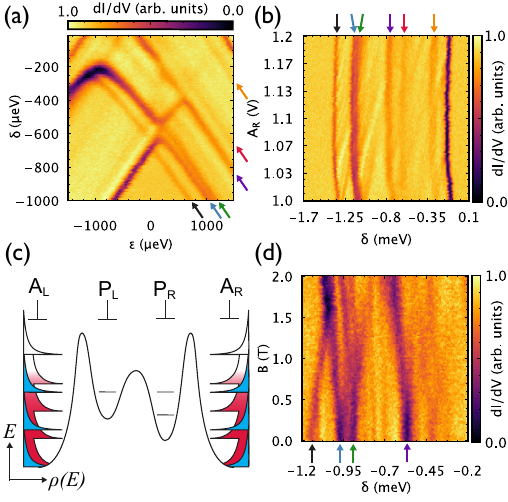}
\caption{\bf Identifying relevant dot states.  \rm (a) Overlaid DAXS data with colored arrows pointing to states corresponding to the right quantum dot.  (b) Plotting $\delta$ vs.~$A_\text{R}$ with pulsed gate voltages applied.  Vertical lines with colored arrows correspond to dot states while diagonal lines correspond to resonances in the leads.  Compensation is applied to the $P_\text{R}$ as well as $B_\text{R}$ voltages to keep the lines of dot states vertical and hold tunnel rates relatively constant, respectively.  (c) Schematic depicting a picture of the Fermi level of the leads and the resonances due to the quasi-1D behavior~\cite{mottonenProbeControlReservoir2010}.  Red and blue correspond to how lead states may look at different lead voltages. (d) A combined magnetospectroscopy and PGS scan used to examine the behavior of excited singlet and triplet states in a magnetic field.}
\label{fig4}
\end{figure}

The colored arrows in Fig.~\ref{fig4}(a) highlight peaks from the data shown in Fig.~\ref{fig12}(d) that we argue arise from the DQD and not from the reservoirs. To make this argument, we perform a series of measurements at $\varepsilon = 1.5$~meV.  Fig.~\ref{fig4}(b) shows repeated DAXS measurement at that detuning for many different right accumulation gate voltages $A_\text{R}$. During the measurement to mitigate crosstalk, a compensation voltage is applied to the dot plunger gates
and to the tunnel barrier gates. In the resulting color plot of Fig.~\ref{fig4}(b), many lines shift rapidly as a function of the voltage on $A_\text{R}$. Because $A_\text{R}$ controls the energy of the quasi-1D resonances in the reservoir, as shown in Fig.~\ref{fig4}(c), we attribute these sloped lines as corresponding to those resonances. In contrast, six of the lines in Fig.~\ref{fig4}(b) are nearly vertical in that plot; these are marked with arrows of the same color as in panel (a), and they are independent of the reservoir gate voltage, indicating that they arise from the DQD and not from quasi-1D resonances in the reservoirs.

\textit{Magnetospectroscopy}---In Fig.~\ref{fig4}(d), we report a measurement analogous to that in Fig.~\ref{fig4}(b) in which we vary the in-plane magnetic field and hold the reservoir gate voltages constant. As expected, the lines corresponding to energy eigenstates display different behavior as a function of $B$, and we argue that these shifts arise from changes in energy corresponding to the Zeeman effect. The lowest lying state, which is highlighted with a black arrow, does not split consistent with the two-electron ground state singlet primarily localized in the right quantum dot. The next state, in blue, corresponds to the first excited state, which is expected to be a triplet state; indeed, this line splits into 3 lines as $B$ is increased.  The third state, labeled with a green arrow, behaves similarly to the ground state, providing evidence that it too is a spin singlet. Finally, the state labeled by the purple arrow is treated as a triplet in the fitting described above because it also appears to split in response to the Zeeman effect.
%

%
\textit{Conclusion and Outlook}---We have shown that DAXS is a simple yet powerful technique that utilizes low-frequency baseband pulsing to extract parameters in the DQD Hamiltonian matrix.  Using energy-calibration lever arms, DAXS plots can be fit to the Hamitonian eigenvalues to extract both the on and off-diagonal elements of the Hamiltonian.  Here we demonstrated DAXS in the (1,3)-(0,4) electron regime, but it can be used in any configuration in which the energy spacings are greater than the temperature of the electron reservoirs.  Because DAXS can be utilized to extract several tunnel couplings at once, it could be used to probe the relationships between ground and excited couplings within DQD systems.  Furthermore, while DAXS requires the continuous loading and unloading of a DQD system, it does not necessarily require reservoirs. Future work could explore using neighboring quantum dots inside large arrays in place of reservoirs as the source of such charge carriers.

\textit{Acknowledgements}---The authors acknowledge Owen Eskandari for useful discussions.  This material is primarily supported by the U.S. Department of Energy, Office of Science, National Quantum Information Science Research Centers as part of the Q-NEXT center. We acknowledge support from Intel Corporation under Cooperative Agreement No. W911NF-22-2-0037 for providing the device studied here and for useful discussions including with Joelle Corrigan, Matthew Curry, and Nathaniel Bishop.  This research was sponsored in part by the Army Research Office under Award No. W911NF-23-1-0110 for sample mount fabrication and fabrication of an initial test device, and under W911NF-18-1-0106 for aiding in integration of electronics. We also thank HRL Laboratories, LLC for support useful in the measurement setup.  The views, conclusions, and recommendations contained in this document are those of the authors and are not necessarily endorsed by nor should they be interpreted as representing the official policies, either expressed or implied, of the Army Research Office or the U.S. Government. The U.S. Government is authorized to reproduce and distribute reprints for U.S. Government purposes notwithstanding any copyright notation herein.

\clearpage

\onecolumngrid

\setcounter{section}{0}
\renewcommand{\thesection}{S\arabic{section}}

\setcounter{figure}{0}
\renewcommand{\thefigure}{S\arabic{figure}}
\renewcommand{\figurename}{Figure} 

\setcounter{equation}{0}
\renewcommand{\theequation}{S\arabic{equation}}

\setcounter{table}{0}
\renewcommand{\thetable}{S\arabic{table}}

\section*{Supplementary Material}

\section{S1. Experimental Setup}
\label{sec:si-1}

In this experiment, the device is mounted at the mixing plate of a dilution refrigerator with a base temperature of 7 mK.  DC control of voltages on the gates of the device is performed with SIM 928 floating voltage sources in SIM900 mainframes at room temperature passed down to the device through cryogenic loom.  Additionally, two separate lock-in amplifiers couple through the loom to the plunger gates on the device at two separate frequencies.  High frequency (MHz bandwidth) is used on plunger gates $P_\text{L}$ and $P_\text{R}$.  The high frequency and DC components are combined at bias tees on the device PCB.  Readout is performed by measuring the change in current through the charge-sensing quantum dot.  This current is amplified at room temperature through a transimpedance,  amplifier (DL1211) and then through a differential voltage amplifier (SR-560).  Finally, the signal is digitized at a computer through an NIDAQ as well as lock-in ADCs. The separate lock-in measurements corresponding to each plunger gate are normalized and added together for each dataset.  


\section{S2. Fitting Process} \label{sec:fitting}
\label{sec:si-2}
\begin{figure}
\centering
\includegraphics{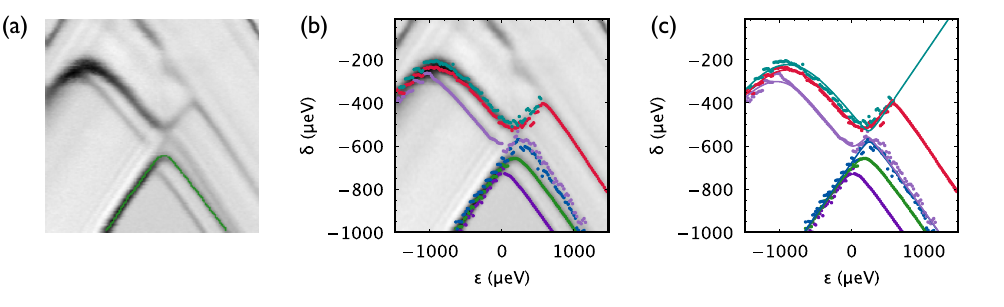}
\caption{\bf Steps of the DAXS Fitting Procedure. \rm (a) Example hand-drawn curve used to produce initial guess values of $\delta$ for the Lorentzian fits of each column of data in the image. Similar curves are drawn over each of the other visible eigenstates that are fitted to. (b) Calculated peak locations from the Lorentzian fitting procedure. Each color corresponds to a unique eigenstate. (c) Calculated eigenstates (solid lines) from fitting the Hamiltonian matrix to the Lorentzian data (colored points) from (b).}
\label{figS1}
\end{figure}

Fig.~\ref{figS1} outlines the multi-step fitting process followed to process the DAXS data. First, a GUI is used to hand-draw curves over each of the visible energy states in the image, as demonstrated in Fig.~\ref{figS1}(a). As an additional preprocessing step, a Savitzky-Golay filter is applied along the vertical axis of the image. This filter works by fitting low-order polynomials within a sliding window and applying the resulting coefficients as a convolution, which acts to reduce noise while preserving peak shapes. Then, for each column of data in the image, Lorentzian functions are fit to the data with a nonlinear least-squares fit. Lorentzians are used due to their simplicity. The $\delta$ values from the hand-drawn curves at that particular $\varepsilon$ are used as guess values for the centers of each of the Lorentzian peaks, as shown in Fig.~\ref{figS1}(b). The widths of the Lorentzian peaks are also bounded during the fitting process to ensure the fit is not overly sensitive to noise. The calculated values for the centers of the Lorentzian peaks are then treated as experimental measurements the Hamiltonian eigenvalues for that particular $\varepsilon$. This procedure yields a one-dimensional dataset of the energy of of the Hamiltonian eigenstates as a function of $\varepsilon$.

\begin{figure}
\centering
\includegraphics{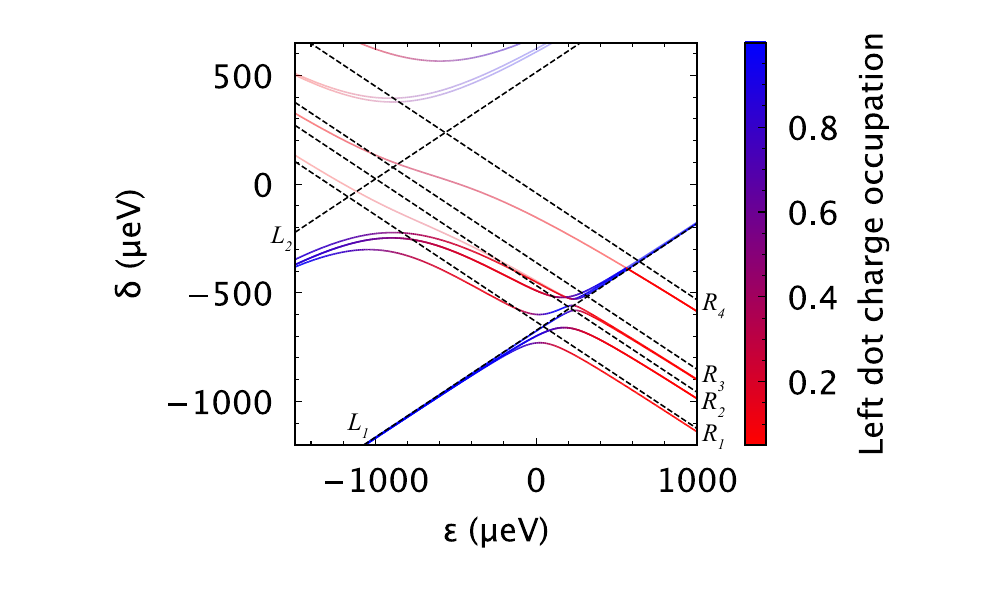}
\caption{\bf Plot of Hamiltonian eigenvalues. \rm Eigenvalues from the fit shown in Fig.~\ref{figS1}, with the range extended. State curves that are not included in the fitting process are shown as partially transparent, and states where the electron is loaded purely into the left/right dot are denoted with dashed black lines.}
\label{figS2}
\end{figure}

Fig.~\ref{figS1}(c) shows the calculated eigenvalues using the fit of the Hamiltonian matrix to these one dimensional datasets. This fit too is performed using a nonlinear least-squares approach, where the fitting function is defined as the concatenation of each of the numerically calculated eigenvalue curves of the Hamiltonian. The Hamiltonian itself is a 15x15 matrix consisting of uncoupled triplet and singlet sub-matrices, given in the block form shown in Eq.~\ref{eq:S1}. The individual triplet and singlet sub-matrices are given in Eq.~\ref{eq:S2} and Eq.~\ref{eq:S3}, respectively. Because the valley splitting is too low to measure with pulsed-gate techniques, we make the simplifying assumption that the valley phase between the two dots for any combination of orbital states is zero. As a result, the triplet sub-matrix can be further decomposed into two identical, uncoupled matrices, one for each set of valley states. This is not true of the singlet sub-matrix, however, because the unloaded right dot contains three electrons, meaning that only one valley state is accessible when a fourth electron loads into the right dot ground state singlet. States where an electron is loaded primarily into the left (right) dot are denoted $L_n$ ($R_n$), with n being a numbering of the visible states in that dot, starting from ground. Duplicate valley states are delineated with $_{V,G}$. $R_1$ and $R_3$ states are singlets, while $R_2$ and $R_4$ states are triplets. A full plot of the eigenvalues along with the labeled basis states is given in Fig.~\ref{figS2}.

\FloatBarrier
\begin{equation}
H =
\vcenter{\hbox{$
\begin{bNiceMatrix}[columns-width=auto]
H_T & \mathbf{0} \\
\mathbf{0} & H_S
\end{bNiceMatrix}
$}}
\label{eq:S1}
\end{equation}

\begin{equation}
\resizebox{0.8\textwidth}{!}{
$
H_T =
\begin{NiceMatrix}[columns-width=auto,baseline=5]
 & L_{2V} & L_{2G} & L_{1V} & L_{1G} & R_{4V} & R_{4G} & R_{2V} & R_{2G}\\
    L_{2V} & \frac{\varepsilon}{2} + l_{2 1} & 0 & 0 & 0 & t_{4 2} & 0 & t_{2 2} & 0 \\
    L_{2G} & 0 & \frac{\varepsilon}{2} + l_{2 1} & 0 & 0 & 0 & t_{4 2} & 0 & t_{2 2} \\
    L_{1V} & 0 & 0 & \frac{\varepsilon}{2} & 0 & t_{4 1} & 0 & t_{2 1} & 0 \\
    L_{1G} & 0 & 0 & 0 & \frac{\varepsilon}{2} & 0 & t_{4 1} & 0 & t_{2 1} \\
    R_{4V} & t_{4 2} & 0 & t_{4 1} & 0 & -\frac{\varepsilon}{2} + r_{4 1} & 0 & 0 & 0 \\
    R_{4G} & 0 & t_{4 2} & 0 & t_{4 1} & 0 & -\frac{\varepsilon}{2} + r_{4 1} & 0 & 0 \\
    R_{2V} & t_{2 2} & 0 & t_{2 1} & 0 & 0 & 0 & -\frac{\varepsilon}{2} + r_{2 1} & 0 \\
    R_{2G} & 0 & t_{2 2} & 0 & t_{2 1} & 0 & 0 & 0 & -\frac{\varepsilon}{2} + r_{2 1} \\
\CodeAfter
\SubMatrix[{2-2}{9-9}]
\end{NiceMatrix}
\label{eq:S2}
$
}
\end{equation}

\begin{equation}
\resizebox{0.8\textwidth}{!}{
$
H_S =
\begin{NiceMatrix}[columns-width=auto,baseline=5]
 & L_{2V} & L_{2G} & L_{1V} & L_{1G} & R_{3V} & R_{3G} & R_{1V}\\
    L_{2V} & \frac{\varepsilon}{2} + l_{2 1} & 0 & 0 & 0 & t_{3 2} & 0 & t_{1 2} \\
    L_{2G} & 0 & \frac{\varepsilon}{2} + l_{2 1} & 0 & 0 & 0 & t_{3 2} & 0 \\
    L_{1V} & 0 & 0 & \frac{\varepsilon}{2} & 0 & t_{3 1} & 0 & t_{1 1} \\
    L_{1G} & 0 & 0 & 0 & \frac{\varepsilon}{2} & 0 & t_{3 1} & 0 \\
    R_{3V} & t_{3 2} & 0 & t_{3 1} & 0 & -\frac{\varepsilon}{2} + r_{3 1} & 0 & 0 \\
    R_{3G} & 0 & t_{3 2} & 0 & t_{3 1} & 0 & -\frac{\varepsilon}{2} + r_{3 1} & 0 \\
    R_{1V} & t_{1 2} & 0 & t_{1 1} & 0 & 0 & 0 & -\frac{\varepsilon}{2}
\CodeAfter
\SubMatrix[{2-2}{8-8}]
\end{NiceMatrix}
\label{eq:S3}
$
}
\end{equation}

In order to reduce the visibility of lead resonances in our data, we employ averaging of multiple DAXS scans. We do this by taking a series of DAXS scans at varying reservoir gate voltages. The adjacent barrier gate voltage is also adjusted to maintain a consistent tunnel rate into the double dot across scans. Because the energies of the lead resonances relative to the dot states change with reservoir gate voltage, when these DAXS plots are overlaid the resonant states in the leads get averaged out. In order to ensure proper alignment of the dot states when overlaying, we first fit one branch of a hyperbola to the lower state of the $t_{21}$ anticrossing in each DAXS image. The pixel closest to the vertex of the hyperbola is chosen as the reference point for that image. Each DAXS image is then aligned so that the chosen reference point is placed at the same coordinates.

\FloatBarrier

\section{S3. Determination of coupling signs}
\label{sec:si-3}
It is not immediately clear what the relative signs of all the tunnel couplings should be in the Hamiltonian matrix, because the system includes multi-electron interactions and excited orbital states. However, knowing the exact signs of the tunnel couplings for this specific case is not crucial for demonstrating this method. In the parameter regime measured here, the effect of flipping the sign of any one tunnel coupling on the resulting eigenstates is also relatively small. For these reasons, all fits shown in the main text were performed with all tunnel coupling values constrained to be positive. Nevertheless, for completeness we quantify here the effect the relative tunnel coupling signs have on the resulting fit values. Without direct knowledge of what the signs should be, our approach is to perform separate fits on the data with different unique combinations of tunnel coupling signs enforced. The magnitudes of the fitted tunnel couplings can then be compared between fits, and the difference in each tunnel coupling between fits can simply be included as a known source of systematic error in the estimated fit value.

In the Hamiltonian matrix used here, there are four unique triplet tunnel couplings, and four unique singlet tunnel couplings. For the triplet sub-matrix, flipping the signs of any even number of triplet tunnel couplings is equivalent to simply flipping the signs of basis vectors, so the eigenvalues are unchanged. This means that although there are sixteen possible combinations of signs for the four triplet tunnel couplings, they reduce to two equivalence classes: the first equivalent to all four tunnel coupling signs positive, and the second equivalent to any three tunnel coupling signs positive with the fourth negative. The same argument holds true for the singlet sub-matrix, meaning that it also has only two distinct equivalence classes of tunnel coupling signs that need to be tested. Because the singlet and triplet sub-matrices are not coupled to each other, the span of all tunnel coupling sign combinations for the full Hamiltonian can be tested with only two distinct cases. The first is with all positive tunnel couplings, and the second is with one negative triplet tunnel coupling and one negative singlet tunnel coupling, with the rest of the tunnel couplings positive.

\begin{figure}
\centering
\includegraphics{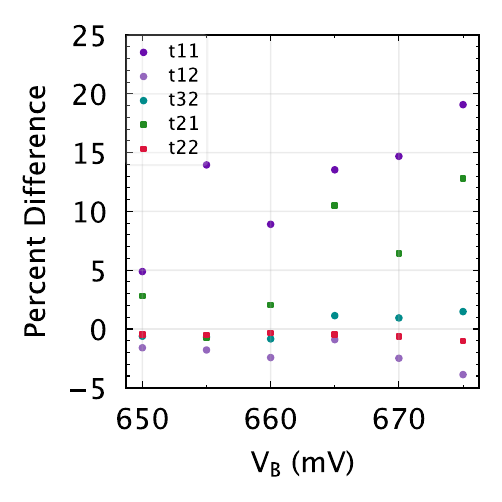}
\caption{\bf Comparison of tunnel coupling fit values between sign combinations. \rm Each data point denotes the percent difference in the magnitude of the given tunnel coupling between the fit performed with all tunnel couplings positive, and the fit performed with one singlet tunnel coupling negative and one triplet tunnel coupling negative. Circular points denote singlet tunnel couplings, and square points denote triplet tunnel couplings. The $t_{31}$, $t_{41}$, and $t_{42}$ tunnel couplings are not included in this plot, for reasons which are discussed in SI Secs.~S4 and S5.}
\label{figS3}
\end{figure}

Fig.~\ref{figS3} shows the percent differences in the fitted tunnel couplings between these two sign combinations. We find that for many tunnel couplings this difference is less than 5\% across the range of center barrier gate voltage values tested. For tunnel couplings $t_{11}$ and $t_{21}$, the percent difference between cases is greater, as much as 20\%, but still comparable to the random error in the fitting procedure caused by charge noise, as discussed in SI Sec.~S5. 

\FloatBarrier
\section{S4. Effect of higher orbital couplings}
\label{sec:si-4}

We recognize that some excited state features are not visible for the tuning and pulse frequency used here. This is because the visibility of excited states in DAXS and similar pulsed-gate techniques depends upon the change in the tunnel rates into the dots. The maximum usable pulse amplitude was also limited in this particular experiment due to degradation of the signal-to-noise ratio for high amplitudes. Because of these limitations, any states above the first excited orbital in the left dot, and any states above the fourth visible state in the right dot, are not included in our fitting model. The anticrossing of the fourth right dot state and the left dot excited orbital state, labeled $t_{4 2}$, is also not visible in our measurements. The inability to directly fit this anticrossing does add uncertainty to the fit of the states that are visible. However, it is not unreasonable for the magnitude of $t_{4 2}$ to be similar to other higher orbital tunnel couplings in the system, so in the fitting procedure we constrain $t_{4 2}$ to be equal to $t_{32}$, the highest orbital coupling that is directly fitted.

\begin{figure}
\centering
\includegraphics{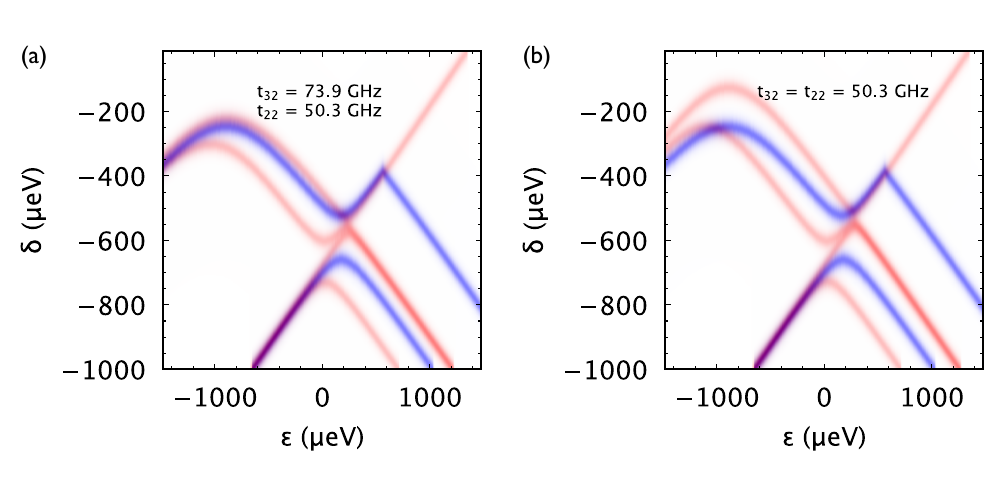}
\caption{\bf Effect of $t_{32}$ coupling on state visibility. \rm (a) Eigenvalues from the fit shown in Fig.~\ref{eq:S1}, plotted with finite linewidths matching the data. Triplet states are colored blue and singlet states are colored red. Note the close proximity of the $R_2$ and $R_3$ states for negative $\varepsilon$. (b) Same plot as (a), but with the value of the $t_{32}$ tunnel coupling in the model decreased to demonstrate its effect on the resulting singlet state energies.}
\label{figS4}
\end{figure}

We also note the apparent disappearance of the $R_3$ singlet state for negative $\varepsilon$  values in the data. We attribute this to a large $t_{32}$ tunnel coupling, which acts to lower the energy of the $R_3$ state to close to the energy of the $R_2$ state, as demonstrated in Fig.~\ref{figS4}. Because of the finite width of the transition lines due to both the electron temperature and tunnel broadening effects, these states become visually indistinguishable near the $t_{32}$ anticrossing. Despite this limitation in visibility, the fitting procedure is still able to extract reasonable estimates for the value of the $t_{32}$ excited state tunnel coupling, as shown in Fig. 2 in the main text. This demonstrates the power of this approach over using a simplified two-level model.

\FloatBarrier
\section{S5. Error Estimation}
\label{sec:si-5}

\begin{figure}
\centering
\includegraphics{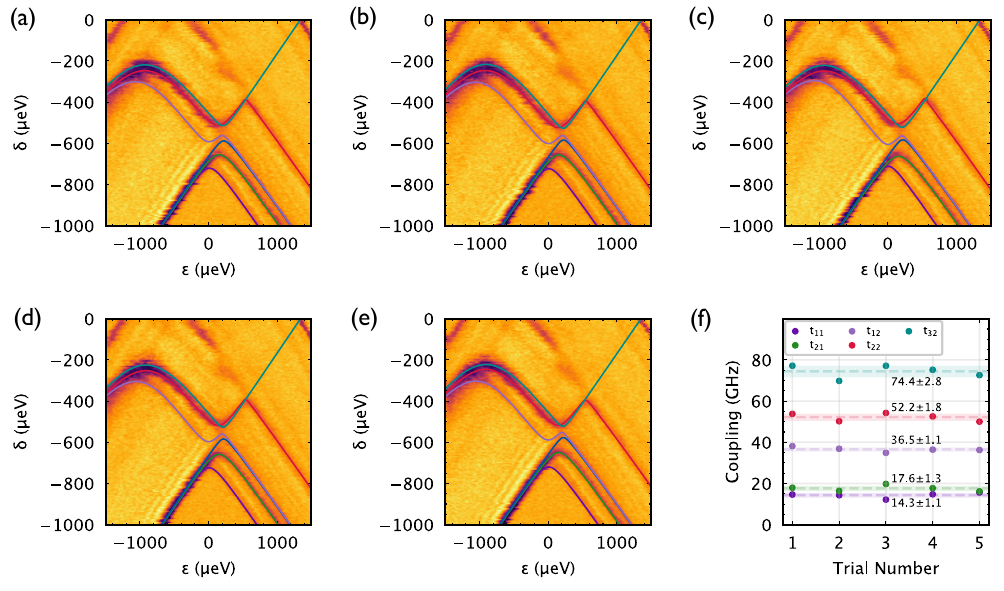}
\caption{\bf Repeated DAXS scan fits. \rm (a)-(e) Five consecutive DAXS scans with the center barrier gate voltage $V_B = 660 mV$, with Hamiltonian state fits overlaid. (f) Extracted tunnel coupling values for each of the five fits. The labels display the mean values $\pm$ one standard deviation. }


\label{figS5}
\end{figure}

To obtain an estimate of the random error in the extracted tunnel couplings due to charge noise, as discussed in the main text we separately take five consecutive DAXS scans at a single device tuning and examine the variability in each of the extracted tunnel couplings across scans, as shown in Fig.~\ref{figS5}. We find that for many of the tunnel couplings in this system, the fitted values remain consistent from scan to scan, with a percent difference between the maximum and minimum values of around 10\%. For two of the tunnel couplings in the system, $t_{31}$ and $t_{41}$, we find a high degree of variability between scans. For the $t_{31}$ tunnel coupling, we attribute this to a lack of direct visibility of the anticrossing in the data, due to the obscuring effect of the nearby $t_{11}$ and $t_{21}$ anticrossings. For the $t_{41}$ tunnel coupling, the value of the coupling is simply too low to accurately measure for the given linewidths. Because of this high variability for these two tunnel couplings, we do not consider the fit values obtained for either to be reliable. Given the number of overlapping states of the system, it is not unexpected that some parameters in the Hamiltonian will not be measurable with great precision, despite the majority displaying strong repeatability between scans. A global multiplicative factor is also included as a fitting parameter for the Hamiltonian eigenvalue fits. This is a correction factor intended to help account for any error in the lever arm estimates used to convert from gate voltages to $\varepsilon$ and $\delta$. In all fits performed, this value is within 4\% of unity.

\FloatBarrier
\section{S6. Additional DAXS Plots}

\begin{figure}
\centering
\includegraphics{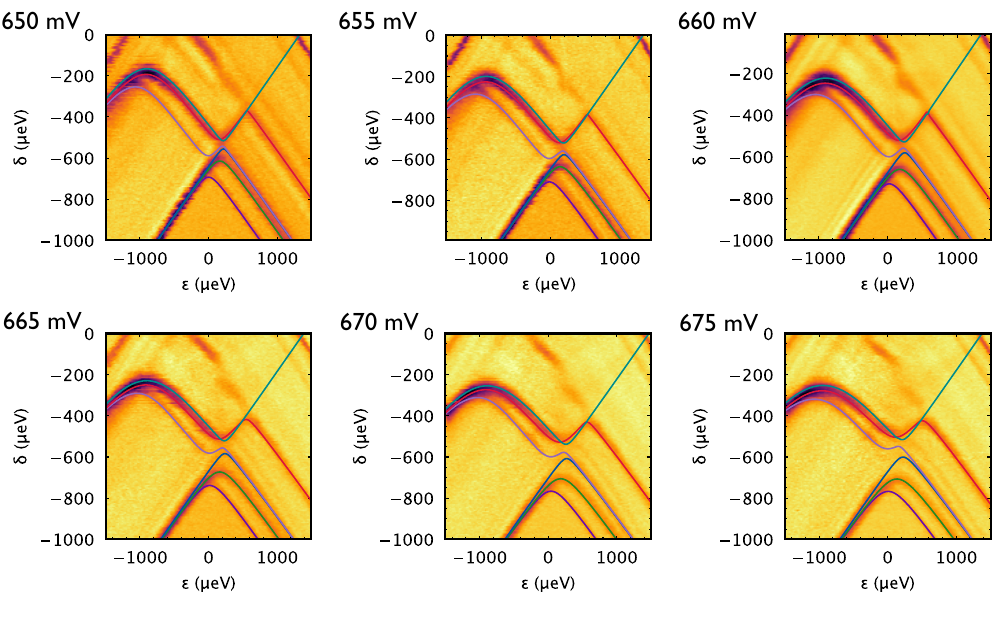}
\caption{\bf DAXS measurements at varying center barrier gate voltages. \rm DAXS datasets for the series of center barrier gate voltages shown in Fig. 2(b), with the eigenvalue fits overlaid.}
\label{figS6}
\end{figure}

For completeness we display in Fig.~\ref{figS6} the fitted DAXS datasets at varying $B_C$ gate voltages used to make Fig. 2(b).

\FloatBarrier
\section{S7. Conversion from gate voltages to double quantum dot energies}

To convert from plunger gate voltages $P_L$ and $P_R$ to the double quantum dot parameters detuning $\varepsilon = \mu_2 - \mu_1$ and delta $\delta = \frac{\mu_1 + \mu_2}{2}$ (also defined in the main text), we first virtualize the plunger gate voltages with respect to each other using Eq.~\ref{eq:S4}, with $\alpha_n^m$ denoting the lever arm from gate $n$ to quantum dot $m$. The virtual plunger gate voltages $P_{Lv}$ and $P_{Rv}$ can then be converted to chemical potentials $\mu_1$ and $\mu_2$ by multiplying each by their respective lever arms, $\alpha_2^2$ and $\alpha_3^3$. From there $\varepsilon$ and $\delta$ can be calculated using their given definitions.

\begin{equation}
\begin{aligned}
P_{Lv} = P_L + \frac{\alpha_3^2}{\alpha_2^2}P_R
&\qquad
P_{Rv} = P_R + \frac{\alpha_2^3}{\alpha_3^3}P_L
\end{aligned}
\label{eq:S4}
\end{equation}

\end{document}